# Epicenter localization using forward-transmission laser interferometry

BOHAN ZHANG,[1] GUAN WANG,[1,2] ZHONGWANG PANG,[1,2] AND BO WANG[1,2,*]

[1] *State Key Laboratory of Precision Measurement Technology and Instruments, Department of Precision Instrument, Tsinghua University, Beijing 100084, China*

[2] *Key Laboratory of Photonic Control Technology (Tsinghua University), Ministry of Education, Beijing 100084, China*

*\* [bo.wang@tsinghua.edu.cn](bo.wang@tsinghua.edu.cn)*

**Abstract:** Widely distributed optical fibers, together with phase-sensitive laser interferometry, can expand seismic detection methods and have great potential for epicenter localization. In this paper, we propose an integral response method based on a forward transmission scheme. It uses spectrum analysis and parameter fitting to localize the epicenter. With the given shape of the fiber ring, the integral phase changes of light propagating in the forward and reverse directions can be used to determine the direction, depth, distance of the epicenter, and seismic wave speed. For the noisy case with SNR=20 dB, the simulation results show ultrahigh precision when epicenter distance is 200 km: the error of the orientation angle is ~0.003°±0.002°, the error of the P-wave speed is ~0.9±1.2 m/s, the error of the epicenter depth is ~9.5±400 m, and the error of the epicenter distance is ~200±760 m.

## 1. Introduction

Forward transmission [1] and backscattering [2–11] are two main schemes of fiber vibration sensing technology. The backscattering scheme can use the discrete Direction of Arrival (DOA) technique to find the direction of incoming waves [8,12]. However, it has disadvantages of large optical loss and limited detection bandwidth, which makes it just suitable for dark fibers [4,5]. Furthermore, the low-loss fiber is a never-ending goal of mankind [13-15]. Virtually every Optical Fiber Communication conference features reports of hard-fought battles to eke out even a fraction of a dB/km lower loss each year, because this significantly impacts system reach and cost over long-haul distances [16]. However, these hard-fought battles are not good news for backscattering laser interferometry, which relies on Rayleigh scattering. The forward transmission scheme overcomes the above problems of backscattering. However, it has drawbacks in vibration localization. Since the forward transmission scheme detects the phase change of whole fiber, how to analyze vibrations occurring in different locations is the main problem.

Recently, the preliminary wave method was proposed to localize epicenters [1]. As shown in Fig. 1(a), on each fiber link, the phase difference of two laser interferometers can be obtained, and then points $A_1$ and $A_2$ where the seismic wave first arrives at link 1 and link 2 can be determined. The normal line of points $A_1$ and $A_2$ will point to the epicenter. When two fiber links are implemented, the intersection of two normal lines is the epicenter.

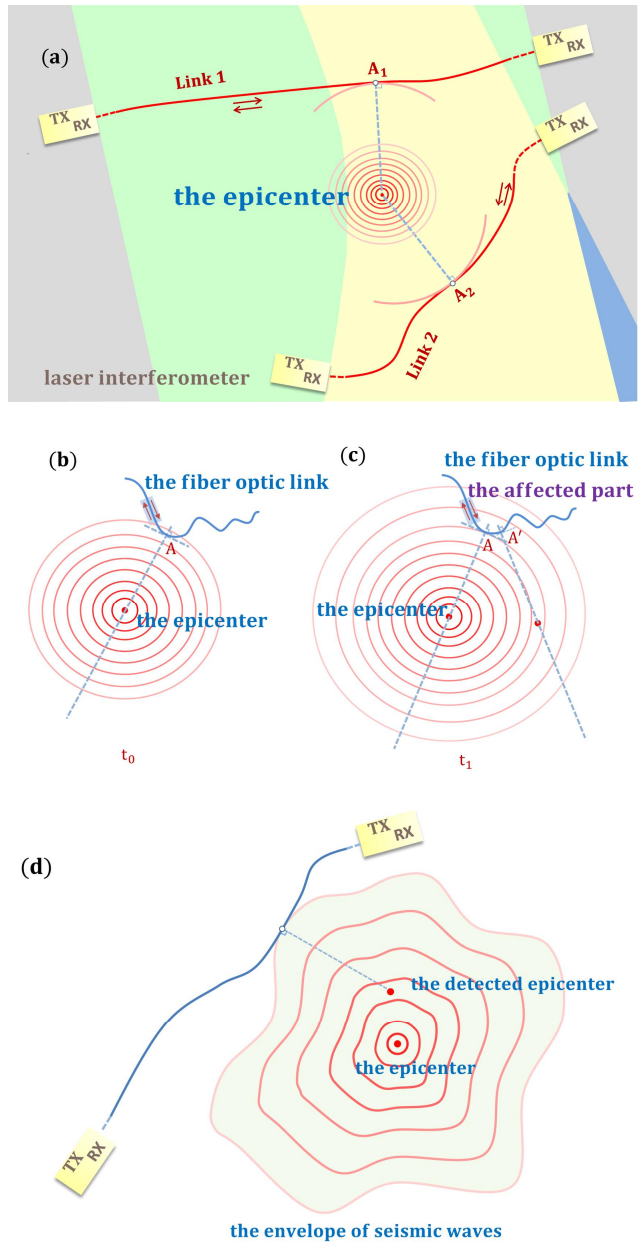

**Fig. 1.** Schematic diagrams of the preliminary wave method and its defects. (a) Schematic diagram of the preliminary wave method and the detection blind area of two fiber links. (b) The schematic diagram when the seismic wave first arrives. (c) The schematic diagram when the derived preliminary point A' deviates from the real point A. (d) Schematic diagram of seismic envelope deformation.

There are several potential problems with this method. First, since seismic vibration is a nonstationary process, the intensity envelope of seismic waves gradually becomes stronger. This means that the strength of the preliminary wave is weak, and the affected fiber length is limited (Fig. 1(b)). This will lead to a low signal-to-noise ratio (SNR), which dramatically affects the localization result [17]. Second, with the propagation of the seismic wave (Fig. 1(c)), a larger section of fiber will be disturbed by the earthquake. The calculated point A' determined by the preliminary wave method will deviate from the real point A, and the detected direction of the epicenter will deviate from the real direction. Third, for a certain fiber layout, vibration localization will have a large blind area using the preliminary wave method. For example, when the end point of a fiber link is the nearest point of the link to the epicenter, this method will fail to localize it. Fig. 1(a) shows the blind area of the fiber link layout in Ref. [1]. The yellow part is the locatable area for both links. The green and blue parts are the locatable areas for just one link. The gray part indicates the detection blind area.

Furthermore, the seismic envelope consists of several seismic waves caused by different propagation paths, and the envelope may be formed by the wavefront in different directions. The prerequisite of the preliminary wave method is that the propagation speeds of the seismic waves are the same along different directions. Actually, this is usually the case when seismic waves propagate in the upper crust (epicenter distance less than 200 km and epicenter depth less than 15 km). In this case, the preliminary wave propagates directly from the epicenter to the ground, referred to as direct wave. With the increase of epicentral distance and depth, other seismic waves may overtake direct wave. In this case, the fastest wave is head wave, which travels through the mantle with faster speed than that in upper crust. Therefore, the seismic envelope is complex due to different crust thickness and seismic wave velocity in the mantle on a large scale (Fig. 1(d)). This will lead to a distorted seismic envelope and finally cause epicenter positioning errors. Additionally, the preliminary wave method cannot obtain the depth, wave speed and distance of the earthquake event, which limits its broad application.

In this paper, we propose an integral response method based on forward-transmission laser interferometry. The orientation, depth, distance of the epicenter, and seismic wave speed can be determined by integrating the phase changes throughout the fiber link. The simulation result shows that the proposed method can overcome the above 4 problems of the preliminary wave method. For the noisy case with SNR=20 dB, when the epicenter distance is 200 km, the error of the orientation angle is ~0.003°±0.002°, the error of the P-wave speed is ~0.9±1.2 m/s, the error of the epicenter depth is ~9.5±400 m, and the error of the epicenter distance is ~200±760 m.

## 2. Schematic and model

Fig. 2 shows the schematic diagram of epicenter relative to the fiber optic link. $TX_1$, $TX_2$ and $RX_1$, $RX_2$ are the transmitting and receiving ports of two laser interferometers, respectively [18–20]. The phase change of the laser interferometer can be extracted from the interference of the sensing signal with reference signal using a digital IQ demodulator. Considering that the depth of epicenter E is $H$, the projection of epicenter E on the ground is point O, as shown in Fig. 2 (a). To analyze the earthquake effect on the ground, a cylindrical coordinate system is established with O as the origin and the vertical direction as the cylindrical axis z. Any horizontal direction can be set as the polar axis x. Considering the microelement $dl$ on the fiber link at point O', its position can be expressed using polar diameter *r*, angular position $\theta$ and height *z*, where height *z*=0. The direction vectors are the radial direction vector $\hat{r}$, the circumferential direction vector $\hat{\theta}$, and the vertical direction vector $\hat{z}$. The fiber length from the link's end point to microelement *dl* is *l* (clockwise direction). The angle between the tangent direction and the radial direction $\hat{r}$ on the microelement is *β*. The incident angle between the wave vector $\hat{k}$ and the vertical direction $\hat{z}$ is $i_p$. The distance from the centroid C of the fiber link to the epicenter E is *D*. Fig. 2(b) shows the top view of Fig. 2(a). The radial direction vector at point C is $\hat{r}'$. To describe the shape of the fiber optic link and its location relative to the epicenter E, we establish a polar coordinate system with C as the pole, and the direction from the observation station to the pole is the reference direction of the polar axis x'. The polar angle of O in the polar coordinate system is $\alpha$.

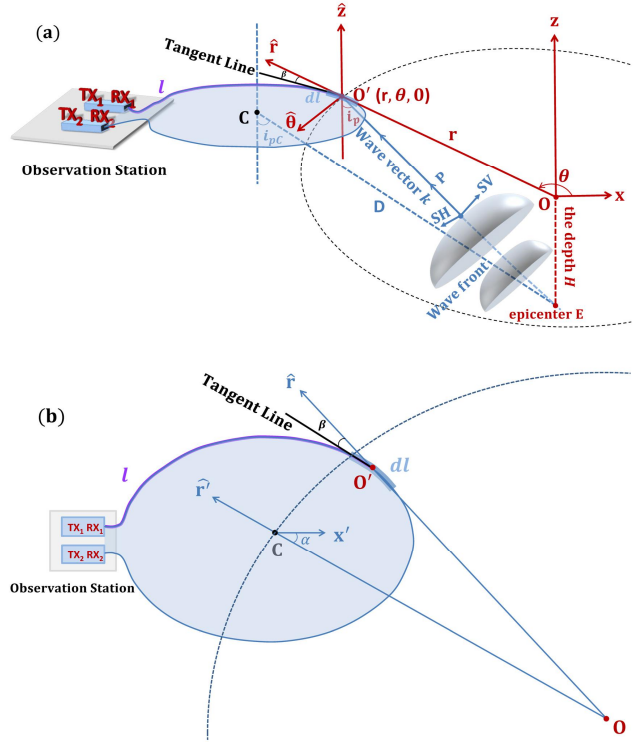

**Fig. 2.** (a) Schematic diagram of the epicenter relative to the fiber optic link. (b) Top view of the schematic diagram.

To calculate the phase change of the laser interferometer, a model should be built to describe the earthquake-induced strain throughout the fiber link. Seismic waves can be classified into P-wave, SH-wave and SV-wave. Normally, P-wave comes first. Taking the P-wave as an example, the displacement induced by the P-wave is only in the $\hat{r}$-$\hat{z}$ plane. First, we analyze the radial displacement of centroid C in the frequency domain, which can be expressed as $u_{C,r}(\omega)$, where $\omega$ is the frequency of the seismic wave. Then, we analyze the earthquake effect on the micro element *dl*. The radial displacement on the micro element *dl* can be expressed as $u_{dl,r}(\omega)$ [21]:

$$u_{dl\text{-}r}(\omega) = \frac{\eta D}{d} e^{-ik(d-D)} u_{C\text{-}r}(\omega) \tag{1}$$

where $d$ is the distance from point O′ to the epicenter E and can be obtained from the geometric relationship, $\eta$ is the horizontal displacement ratio between points C and O′, which can be determined by different incident angles ($i_{pC}$ and $i_p$) of the seismic wave. A detailed analysis of the parameter $\eta$ is given in the supplementary material. In Eq. 1, we consider the variation of amplitude caused by different distances between the micro element $dl$ and the epicenter E, and the phase delay caused by the arrival time difference between the micro element $dl$ and the centroid C. According to the strain-displacement relations in curvilinear cylindrical coordinates [22], the radial linear strain $\varepsilon_{dl\text{-}r}(\omega)$ and circumferential linear strain $\varepsilon_{dl\text{-}\theta}(\omega)$ together with the shear strain $\varepsilon_{dl\text{-}r\theta}(\omega)$ of micro element $dl$ can be deduced from the radial displacement of centroid C:

$$\varepsilon_{dl\text{-}r}(\omega) = \frac{\partial u_{dl\text{-}r}(\omega)}{\partial r} = \frac{\partial}{\partial r}[\frac{\eta D}{d} e^{-ik(d-D)}] u_{C\text{-}r}(\omega), \tag{2}$$

$$\varepsilon_{dl\text{-}\theta}(\omega) = \frac{1}{\sqrt{d^2-H^2}} u_{dl\text{-}r}(\omega) = \frac{\eta D}{d\sqrt{d^2-H^2}} e^{-ik(d-D)} u_{C\text{-}r}(\omega), \tag{3}$$

$$\varepsilon_{dl\text{-}r\theta}(\omega) = 0 \tag{4}$$

The fiber in the horizontal plane is insensitive to the shear strain and the linear strain perpendicular to the tangential direction of the fiber [6,23]. Therefore, we can deduce the strain parallel to the tangential direction of the fiber $\varepsilon_{dl}(\omega)$ from applying Eq. 2~4 to strain transformation model [24]:

$$\varepsilon_{dl}(\omega) = \varepsilon_{dl\text{-}r}(\omega)\cos^2\theta + \varepsilon_{dl\text{-}\theta}(\omega)\sin^2\theta + 2\varepsilon_{dl\text{-}r\theta}(\omega)\sin\theta\cos\theta$$
$$= \frac{\partial}{\partial r}[\frac{\eta D}{d} e^{-ik(d-D)}] u_{C\text{-}r}(\omega)\cos^2\theta + \frac{\eta D}{d\sqrt{d^2-H^2}} e^{-ik(d-D)} u_{C\text{-}r}(\omega)\sin^2\theta, \tag{5}$$

where $k$ is the quantity of the P-wave vector $\vec{k}$ with the value of $\frac{\omega}{v_p}$, and $v_p$ is the P-wave speed. It's reasonable to assume that the fiber is well coupled to the cable, so that the tangential strain of the fiber is equal to that of the cable [6,7,25]. Besides, since the Young's modulus of cable is much smaller than that of fiber, the transversal strain of the fiber caused by that of the cable can be neglected [25]. Thus, the phase change $d\varphi$ caused by the P-wave is proportional to the tangential strain on micro element $dl$ [6,25]:

$$d\varphi = \xi \varepsilon_{dl}(\omega) dl \tag{6}$$

where $\xi$ is the phase change of the unit length caused by the unit linear strain. It is dependent on the physical properties of the fiber, such as the refractive index, Pockels constant, and central wavelength of the laser.

## 3. Method and Results

In the proposed integral response scheme, the phase change of two laser interferometers can be calculated theoretically as follows:

$$\varphi^+(\omega) = \int_L a_{dl}(\omega) e^{-i\omega \frac{n(L-l)}{c}} dl$$
$$= \int_L \alpha \{\frac{\Delta}{\Delta r}[\frac{\Delta D}{d} e^{ik(d-D)}]\cos^2\theta + \frac{\Delta D}{d\sqrt{d^2+H^2}} e^{ik(d-D)} \sin^2\theta\} u_{C-r}(\omega) e^{-i\omega \frac{n(L-l)}{c}} dl,  \quad (7)$$

$$\varphi^-(\omega) = \int_L a_{dl}(\omega) e^{-i\omega \frac{nl}{c}} dl$$
$$= \int_L \alpha \{\frac{\Delta}{\Delta r}[\frac{\Delta D}{d} e^{ik(d-D)}]\cos^2\theta + \frac{\Delta D}{d\sqrt{d^2+H^2}} e^{ik(d-D)} \sin^2\theta\} u_{C-r}(\omega) e^{-i\omega \frac{nl}{c}} dl,  \quad (8)$$

where $L$ is the total length of the fiber, $\varphi^+(\omega)$ and $\varphi^-(\omega)$ are the spectra of interference signals clockwise and counterclockwise in the top view, respectively. The phase offsets $-\frac{n(L-l)}{c}$ and $-\frac{nl}{c}$ are caused by different phase time delays from micro element $dl$ to RX1 and RX2, respectively. Here, $n$ is the refractive index of the fiber, and $c$ is the vacuum speed of light.

To normalize the amplitudes, we divide $\varphi^+(\omega)$ by $\varphi^-(\omega)$ and obtain the Link Response Function (LRF):

$$LRF(\alpha, v_p, D, H, \omega) = \frac{\varphi^+(\omega)}{\varphi^-(\omega)} = \frac{\int_L \{\frac{\Delta}{\Delta r}[\frac{\Delta}{d} e^{ik(d-D)}]\cos^2\theta + \frac{\Delta}{d\sqrt{d^2+H^2}} e^{ik(d-D)} \sin^2\theta\} e^{-i\omega \frac{n(L-l)}{c}} dl}{\int_L \{\frac{\Delta}{\Delta r}[\frac{\Delta}{d} e^{ik(d-D)}]\cos^2\theta + \frac{\Delta}{d\sqrt{d^2+H^2}} e^{ik(d-D)} \sin^2\theta\} e^{-i\omega \frac{nl}{c}} dl}.$$
(9)

In Eq. 9, the terms $\alpha$, $D$ and $u_{C-r}(\omega)$ have been canceled because they have no relation with the variable $dl$. LRF is a function of the seismic frequency $\omega$, the epicentral direction $\alpha$, the P-wave speed $v_p$, the distance from the centroid to epicenter $D$, and the depth of epicenter $H$. Actually, the terms $H$ and $\alpha$ are hidden in variables $\theta$, $\phi$ and $d$.

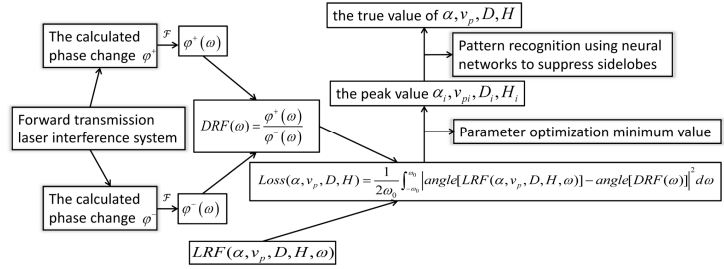

**Fig. 3.** Flow chart of the proposed method.

Using the forward-transmission laser interferometry, the phase change $\varphi^+(t)$ and $\varphi^-(t)$ can be detected. Through Fourier transform, we can obtain their spectra $\varphi^+(\omega)$ and $\varphi^-(\omega)$, respectively. $DRF(\omega) = \varphi^+(\omega)/\varphi^-(\omega)$ is the Detected Response Function (DRF). According to the minimum mean square error principle (MMSE), we define the loss function $Loss(\alpha, v_p, D, H)$ as:

$$Loss(\alpha, v_p, D, H) = \frac{1}{2\omega_0} \int_{-\omega_0}^{\omega_0} \left|angle[LRF(\alpha, v_p, D, H, \omega)] - angle[DRF(\omega)]\right|^2 d\omega. \quad (10)$$

Here $angle[LRF(\alpha, v_p, D, H, \omega)]$ and $angle[DRF(\omega)]$ mean the phase spectrum part of LRF and DRF, respectively. We fit $DRF(\omega)$ with $LRF(\alpha, v_p, D, H, \omega)$ by the genetic algorithm and so on. In other words, we change the value of $\alpha, v_p, D, H$ to minimize the deviation of the LRF from the DRF. Theoretically, the LRF fits the DRF best when $\alpha, v_p, D, H$ is equal to their true value. We can use neural networks to increase the confidence and noise resistance. The detailed methodology of the neural network is shown in the supplementary material. As a summary, we use a flow chart to clearly show steps of the proposed method in Fig. 3.

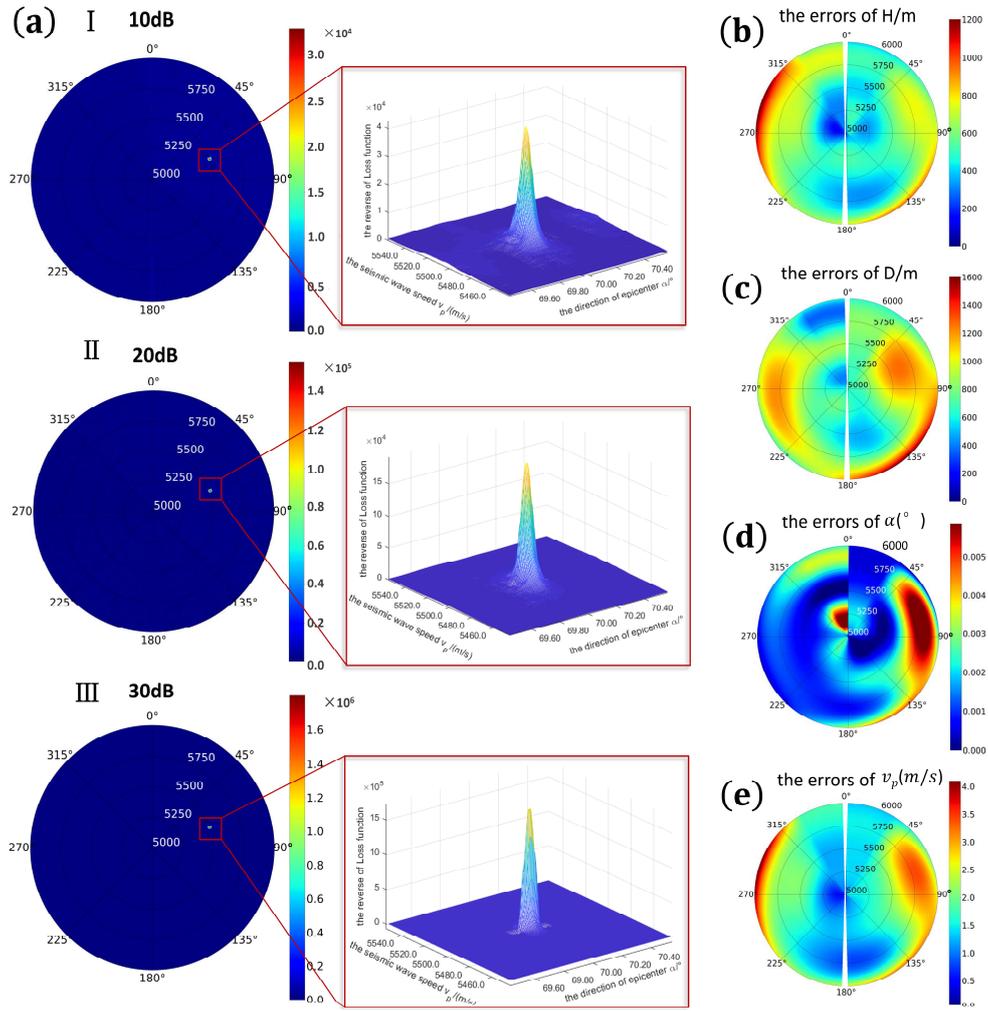

**Fig. 4.** Seismic parameter estimation results of the circle fiber ring under different conditions. (a) Errors of epicenter direction and the wave speed with different SNRs. (b) Errors of epicentral distance $H$ with different epicentral directions and wave speeds. (c) Errors of the epicenter depth $D$ with different epicentral directions and wave speeds. (d) Errors of the epicenter orientation $\alpha$ with different epicentral directions and wave speeds. (e) Errors of the seismic wave speed $v_p$ with different epicentral directions and wave speeds. The polar diameter represents the seismic wave speed $v_p$, and the polar angle is the epicenter orientation $\alpha$.

Table 1. The localization errors of $\theta, v_p, D, H$ with different SNRs

| Fiber ring shape | Circle | Circle | Circle | Square |
|---|---|---|---|---|
| SNR/dB | 10 | 20 | 30 | 20 |
| Bias of $\theta$ /° | 0.0034 | 0.0034 | 8.6×10$^{-5}$ | 0.0033 |
| Halfwidth of $\theta$ /° | 0.011 | 0.0022 | 0.00095 | 0.0013 |
| Bias of $D$/m | 477.4 | 199.7 | 94.7 | 150.0 |
| Halfwidth of $D$/m | 1560 | 760 | 560 | 940 |
| Bias of $H$/m | 18.0 | 9.5 | 3.0 | 1.0 |
| Halfwidth of $H$/m | 700 | 400 | 250 | 350 |
| Bias of $v_p$ /(m/s) | 0.31 | 0.90 | 0.20 | 0.14 |
| Halfwidth of $v_p$ /(m/s) | 2.1 | 1.2 | 1.0 | 0.8 |

To demonstrate the novelty of the proposed integral response method, we carry out a simulation using a circle fiber ring with a radius of 50 km. To improve the credibility of our simulation, we use a real earthquake waveform (event ID: 39338407) from Southern California Earthquake Data Center (SCEDC), to mimic the actual seismic wave [26]. The earthquake event is detected 200 km away from the epicenter, thus the simulation epicentral distance $D$ is set as 200 km. Other parameters are set as follows: the epicenter depth $H$ is 20 km, the P-wave speed is 5500 m/s, and the orientation of epicenter $\theta$ is 70°. We also simulate a 1/f noise to mimic the real noise of detection system [27]. From the detected seismic events of Ref. [1], we can roughly obtain the power ratio between the detected seismic signals and system noise. Thus, we define SNR as the ratio between the power of phase change signal and the power of added 1/f phase noise. Their SNRs are within the range of 10 ~ 30 dB, approximately. Therefore, we carry out simulation with SNR=10 dB, 20 dB and 30 dB to verify the method. The simulation results are shown in Fig. 4 and Table 1. Bias is the deviation from the true value. The halfwidth is the full width at half maximum of the inverse of Loss function in a single measurement. First, we analyze the result of epicenter orientation $\theta$ and P-wave speed $v_p$ under different noise levels. In Fig. 4(a), plots I, II, and III are the localization cases when SNR=10 dB, 20 dB and 30 dB, respectively. They are plotted as the reverse of the loss function. A larger value means a better fitting of the parameters at that point. The polar diameter represents the seismic wave speed $v_p$, and the polar angle is the epicenter orientation $\theta$. The epicenter can be clearly extracted from the polar plot. The errors are given in Table 1. Fig. 4(b), (c), (d) and (e) show the simulation errors of the epicenter depth $H$, the epicenter distance $D$, the epicenter orientation $\theta$ and the seismic wave speed $v_p$ respectively, under different epicenter orientations and P-wave speeds when SNR=20 dB. Because the circle fiber ring used for simulation is symmetric, the error of $v_p$, $H$ and $D$ is larger when the angle between the symmetry axis is within 0.5°, and this part is not included in the figure. This can be avoided by choosing an asymmetric fiber link. When the SNR is 20 dB, the error of the orientation angle is ~0.003°±0.002°, the error of the P-wave speed is ~0.9±1.2 m/s, the error of the epicenter depth is ~9.5±400 m, and the error of the epicenter distance is ~200±760 m.

Actually, the shape of fiber ring doesn't affect the parameter fitting results. As illustrated in Eqs. 7 and 8, the only part related to fiber ring is to integrate the phase change along the fiber. As long as the fiber ring can be described using a coordinate system, it can be taken for arbitrary shape, like the normally existing optical fiber cable. We also carry out a simulation using a square-shaped fiber ring with a side length of 50 km, to demonstrate the practicability of the proposed integral response method. The earthquake parameters are the same as above (plot II of Fig. 4(b)). The simulation results are shown in Fig. 5(a) and Table 1.

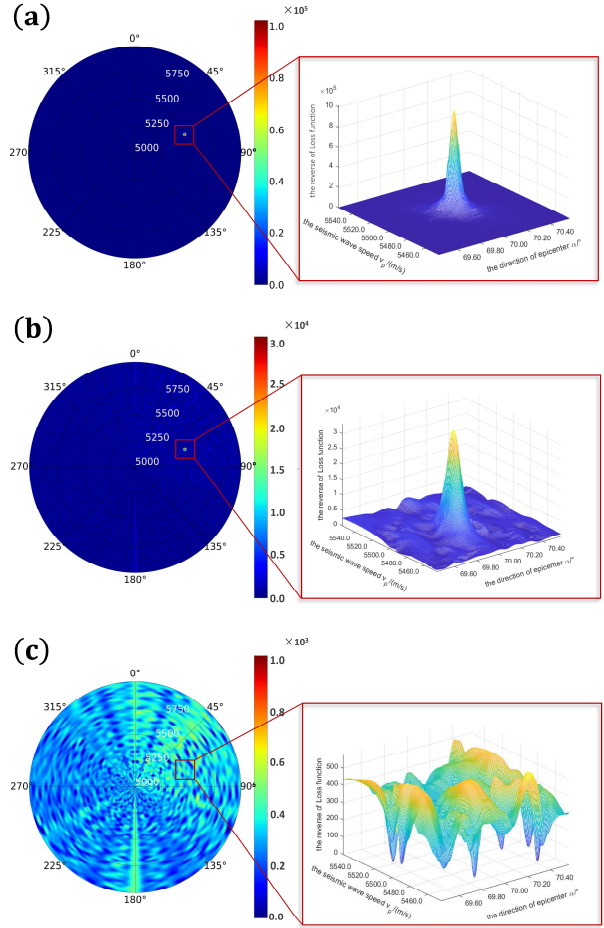

**Fig. 5.** Seismic parameter estimation results under different conditions. (a) Errors of epicenter direction and the wave speed with square-shaped fiber ring. (b) Errors of epicenter direction and the wave speed when affected by both direct waves and head waves. (c) Errors of epicenter direction and the wave speed when affected by head waves only. The polar diameter represents the seismic wave speed $v_p$, and the polar angle is the epicenter orientation $\theta$.

In conclusion, the integral response method can effectively overcome the disadvantages of the preliminary localization method. First, the proposed method is not affected by the low SNR problem of Fig. 1(b). Because it is not limited to detect the preliminary seismic wave, but can select a segment with relatively stronger seismic wave. Second, this method will take the phase changes caused by the seismic wave on the whole fiber into account, rather than only using the detected signal on a certain point. Thus, the deviation between the calculated preliminary point A′ from the actual point A can be avoided. This is demonstrated by the case of the square-shaped fiber ring in Fig. 5(a). Third, the preliminary method is influenced by the envelope distortion caused by the superposition of direct waves and head waves. On the contrary, the proposed integral response method is insensitive to head waves.

To demonstrate that, we carry out simulation under two situations: fiber affected by both direct waves and head waves simultaneously, and fiber only affected by head waves. Considering that the attenuation of head waves is usually much greater than that of direct waves, we set the amplitude ratio between head waves and direct waves to be 1:7. Other earthquake parameters are the same as above and the circle fiber ring is used. The simulation results are shown in Fig. 5(b) and (c). When the fiber is affected by both waves, the epicenter still can be correctly localized, as shown in Fig. 5(b). This is because that the proposed method is insensitive to the head waves. The simulation result of a noisier case can be found in the supplementary material. Fig. 5(c) shows the localization result when the fiber is only affected by the head waves, we won't get a valid localization peak.

Similar conclusions can be made for the SV-wave and SH-wave. The only difference between P-wave and S-wave is: the vibration directions of SV-waves and SH-waves are perpendicular to the direction of wave propagation [21]. For SV-wave, the vibration direction is in the $\hat{r}$-$\hat{z}$ plane of Fig. 2(a). We just need to change the parameter $\alpha$ and replace the wave number of P-wave with the wave number of SV-wave $k_{SV}$, to adjust the model from P-wave case to SV-wave case. $LRF_{SV}$ can be calculated as below:

$$LRF_{SV}(\alpha, v_p, D, H, \omega) = \frac{S_{SV}[\omega[}{S_{SV}[\omega]}$$

$$= \frac{\int_L \{\frac{\partial}{\partial r}[\frac{\eta_{SV}}{d}e^{-ik_{SV}(d-D)}]\cos^2\alpha - \frac{\eta_{SV}}{d\sqrt{d^2+H^2}}e^{-ik_{SV}(d-D)}\sin^2\alpha\}e^{-i\frac{n(L-l)}{c}}dl}{\int_L \{\frac{\partial}{\partial r}[\frac{\eta_{SV}}{d}e^{-ik_{SV}(d-D)}]\cos^2\alpha - \frac{\eta_{SV}}{d\sqrt{d^2+H^2}}e^{-ik_{SV}(d-D)}\sin^2\alpha\}e^{-i\frac{nl}{c}}dl},$$

(11)

where $\eta_{SV}$ is the horizontal displacement ratio between points C and O' induced by the incident SV-wave with the same amplitude. A detailed analysis of the parameter $\eta_{SV}$ is given in the supplementary material. For SH-wave, the vibration direction is in the $\hat{r}$-$\hat{\theta}$ plane of Fig. 2(a). Both $\varepsilon_{dl\text{-}r}(\omega)$ and $\varepsilon_{dl\text{-}z}(\omega)$ are equal to zero while $\varepsilon_{dl\text{-}r\theta}(\omega)$ is not. The $LRF_{SH}$ based on the same model can be easily deduced [22,24]:

$$LRF_{SH}(\alpha, v_p, D, H, \omega) = \frac{S_{SH}[\omega[}{S_{SH}[\omega]}$$

$$= \frac{\int_L \left\{\frac{\partial}{\partial r}\frac{\eta_{SH}}{d}e^{-ik_{SH}(d-D)} - \frac{\eta_{SH}}{d\sqrt{d^2+H^2}}e^{-ik_{SH}(d-D)}\right\}\sin\alpha\cos\alpha \cdot e^{-i\frac{n(L-l)}{c}}dl}{\int_L \left\{\frac{\partial}{\partial r}\frac{\eta_{SH}}{d}e^{-ik_{SH}(d-D)} - \frac{\eta_{SH}}{d\sqrt{d^2+H^2}}e^{-ik_{SH}(d-D)}\right\}\sin\alpha\cos\alpha \cdot e^{-i\frac{nl}{c}}dl}.$$

(12)

where $\eta_{SH} \approx 1$.

Besides, we can use different LRF functions to fit the DRF in different seismic cases. For example, when P-wave is the main component of seismic waves, the correct localization result can be obtained by using the P-wave LRF to fit DRF, and no effective result can be obtained by using $LRF_{SV}$ to fit DRF. When detecting shallow focus earthquakes, the incidence angle is large. $\eta_P$ and $\eta_{SV}$ are much smaller than $\eta_{SH}$ so the effects of P-wave and SV-wave are negligible compared to SH-wave. The phase changes are mainly caused by SH-wave. In this case, we can fit the DRF with $LRF_{SH}$ without being affected by the P-wave and SV-wave.

## 4. Conclusion

In this paper, we analyze the problems of current epicenter localization method. To solve the problems of low SNR, blind zone and localization error of the current scheme, we propose an integral response method. With two counterpropagating laser interferometers, the integral phase changes can be analyzed. From their phase spectra and parameter fitting, the optimal parameters of the epicenter orientation $\theta$, the seismic wave speed, the depth of epicenter $H$ and the distance from centroid to epicenter $D$ can be obtained without relying on a dense seismic network. This method can enrich the earthquake monitoring method and broaden the application fields of existing fiber network.

**Funding.** National Natural Science Foundation of China (62171249, 61971259), and Tsinghua University Initiative Scientific Research Program.

**Disclosures.** The authors declare no conflicts of interest.

**Data availability.** Data underlying the results presented in this paper are not publicly available at this time but may be obtained from the authors upon reasonable request.

**Supplementary Material.** See Supplement 1 for supporting content.